\documentclass[aps,twocolumn,superscriptaddress,showpacs,amssymb,prb,floatfix,longbibliography]{revtex4-1}
\usepackage{graphicx }
\usepackage{dcolumn}
\usepackage{bm}
\usepackage{xcolor}
\usepackage{upgreek}
\usepackage{xspace}
\usepackage{amsmath}
\usepackage[]{hyperref}
\hypersetup{colorlinks=true,linkcolor=blue,citecolor=blue,urlcolor=blue,pdfpagemode=UseNone}

\begin{document}
\newcommand{\naybs}{NaYbS$_2$}
\newcommand{\muSR}{$\mu$SR\xspace}
\newcommand{\mus}{$\upmu$s\textsuperscript{-1}\xspace}
\newcommand{\REF}{{\bf[REF]}}

\title{Quantum spin liquid ground state in the disorder free triangular lattice NaYbS$_2$
}
\author{R. Sarkar}
\email{rajibsarkarsinp@gmail.com}
\affiliation{Institute of Solid State and Materials Physics, TU Dresden, D-01062 Dresden, Germany}
\author{Ph. Schlender}
\affiliation{Faculty of Chemistry and Food Chemistry, TU Dresden, D-01062 Dresden, Germany}
\author{V. Grinenko}
\affiliation{Institute of Solid State and Materials Physics, TU Dresden, D-01062 Dresden, Germany}
\affiliation{Institute for Metallic Materials, IFW Dresden, D-01171 Dresden, Germany}
\author{E. Haeussler}
\affiliation{Faculty of Chemistry and Food Chemistry, TU Dresden, D-01062 Dresden, Germany}
\author{Peter J. Baker}
\affiliation{ISIS Facility, STFC Rutherford Appleton Laboratory, Harwell Oxford, Oxfordshire OX11 0QX, United Kingdom}
\author{Th. Doert}
\affiliation{Faculty of Chemistry and Food Chemistry, TU Dresden, D-01062 Dresden, Germany}
\author{H.-H. Klauss}
\affiliation{Institute of Solid State and Materials Physics, TU Dresden, D-01062 Dresden, Germany}

\date{\today}

\begin{abstract}
Rare-earth delafossites were recently proposed as promising candidates for the realization of an effective $S$\,=\,1/2 quantum spin liquid (QSL) on the triangular lattice. In contrast to the most actively studied triangular-lattice antiferromagnet YbMgGaO$_4$, which is known for considerable structural disorder due to site intermixing, \naybs\ delafossite realizes structurally ideal triangular layers. We present detailed \muSR\ studies on this regular (undistorted) triangular Yb sublattice based system with effective spin $J_{\mathrm{eff}}=1/2$ in the temperature range 0.05 - 40\,K. Zero-field (ZF) and longitudinal field (LF) \muSR\ studies confirm the absence of any long range magnetic order state down to 0.05\,K ($\sim$ $J$/80). 
Current \muSR\ results together with the so far available bulk characterization data suggest that \naybs\ is an ideal candidate to identify QSL ground state.

\end{abstract}

\maketitle
\textit{Introduction.} A quantum spin liquid (QSL) is an exotic state of matter, in which electrons spins are strongly entangled, but do not exhibit any long range magnetic ordering down to $T=0$. Despite considerable effort in the past, so far, experimental realizations of a clean QSL remain scarce. 
At the outset, Anderson proposed that the QSL state can be stabilized in materials where $S\,=\,1/2$ forms  perfect triangular lattice.  
Such a scenario has proved exceptionally hard to realize.~\cite{{Balents-2017},{Balents2010}}
The organic compounds $\kappa$-(BEDT-TTF)$_2$Cu$_2$(CN)$_3$ and EtMe$_3$Sb[Pd(dmit)$_2$]$_2$ appear to be two promising examples of triangular lattices with $S$ = 1/2
moments and fluctuating disordered spin ground states.~\cite{{PhysRevLett.95.036403},{PhysRevB.77.104413}} However, $S = 1/2$ inorganic analogues such as Ba$_3$CoSb$_2$O$_9$  and
NaTiO$_2$ either order magnetically or undergo a lattice deformation on cooling.~\cite{{PhysRevB.76.132407},{PhysRevLett.101.166402},{PhysRevLett.116.087201},{PhysRevLett.108.057205}}

In this context, the triangular lattice magnet YbMgGaO$_4$ was identified as a potential QSL candidate with an effective spin $J_{\mathrm{eff}}$\,=\,1/2.~\cite{Paddison2016} 
YbMgGaO$_4$ contains undistorted triangular planes of magnetic Yb$^{3+}$ with space group $R\bar{3}m$, separated by two triangular planes occupied in a disordered manner by Mg$^{2+}$ and Ga$^{3+}$. The $\mu$SR experiments indicate the absence of static magnetism down to $T$\,=\,50 mK and neutron scattering suggests a continuum of magnetic excitations classifying YbMgGaO$_4$ as hosting a QSL state.~\cite{{PhysRevLett.117.097201},{Li2015},{PhysRevLett.115.167203}} 
However, the influence of Mg$^{2+}$ and Ga$^{3+}$ local disorder on this continuum of magnetic excitations remains the subject of active discussion and study.
   
It was recently proposed that the problem of structural disorder can be overcome in rare-earth delafossites based on Ce or Yb, in which rare-earth ions order into structurally perfect 2D triangular layers. These rare-earth delafossites share the same space group of YbMgGaO$_4$ and the planar triangular spin arrangement.
We have recently reported an extensive study of one of such compounds, \naybs\, using a
combination of thermodynamic, local-probe, and neutron spectroscopy measurements both on high quality single crystals and polycrystalline samples.~\cite{NaYbS2-first-Baenitz} These measurements clearly evidence a strongly anisotropic quasi-2D magnetism and an emerging spin-orbit entangled $S$\,=\,1/2 state of Yb towards low temperatures together with an absence of long-range magnetic order down to 260\,mK. The clear and narrow Yb-ESR lines together with narrow $^{23}$Na NMR lines evidence an absence of inherent structural distortions.
This identifies \naybs\ as a rather pure spin-1/2 triangular lattice magnet and a new candidate quantum spin liquid.~\cite{NaYbS2-first-Baenitz}
\\
\indent
To further investigate \naybs, particularly its nature of the static and/or dynamic ground state, we have performed detailed (\muSR) experiments, both in zero-field (ZF) and in longitudinal field (LF) along the initial muon polarisation, in the temperature range 0.05-40\,K. The main focus was  
in the low temperature region ($T$$\rightarrow$0) and in the longer time window (up to 20\,$\mu$s) to ensure the presence or absence of any long-range static magnetic ordering which is an indispensable information to justify the presence of a QSL ground state. 
Present $\mu$SR studies confirm the absence of any long range magnetic ordering down to 0.05\,K.  
Moreover, in the low temperature limit muon relaxation rates ($\lambda$), which probes the dynamical/static
spin susceptibility at $\mu$eV energy scales, providing vital
information that is complementary to nuclear magnetic resonance and inelastic neutron
scattering, are constant not only for ZF but also for LF=50\,G. This indicates the presence of a highly correlated fluctuating quantum disordered phase in \naybs. Thus, \naybs\ is identified as a candidate material to realize certain class of QSL ground state.

\begin{figure}[h]
\includegraphics[width=\columnwidth]{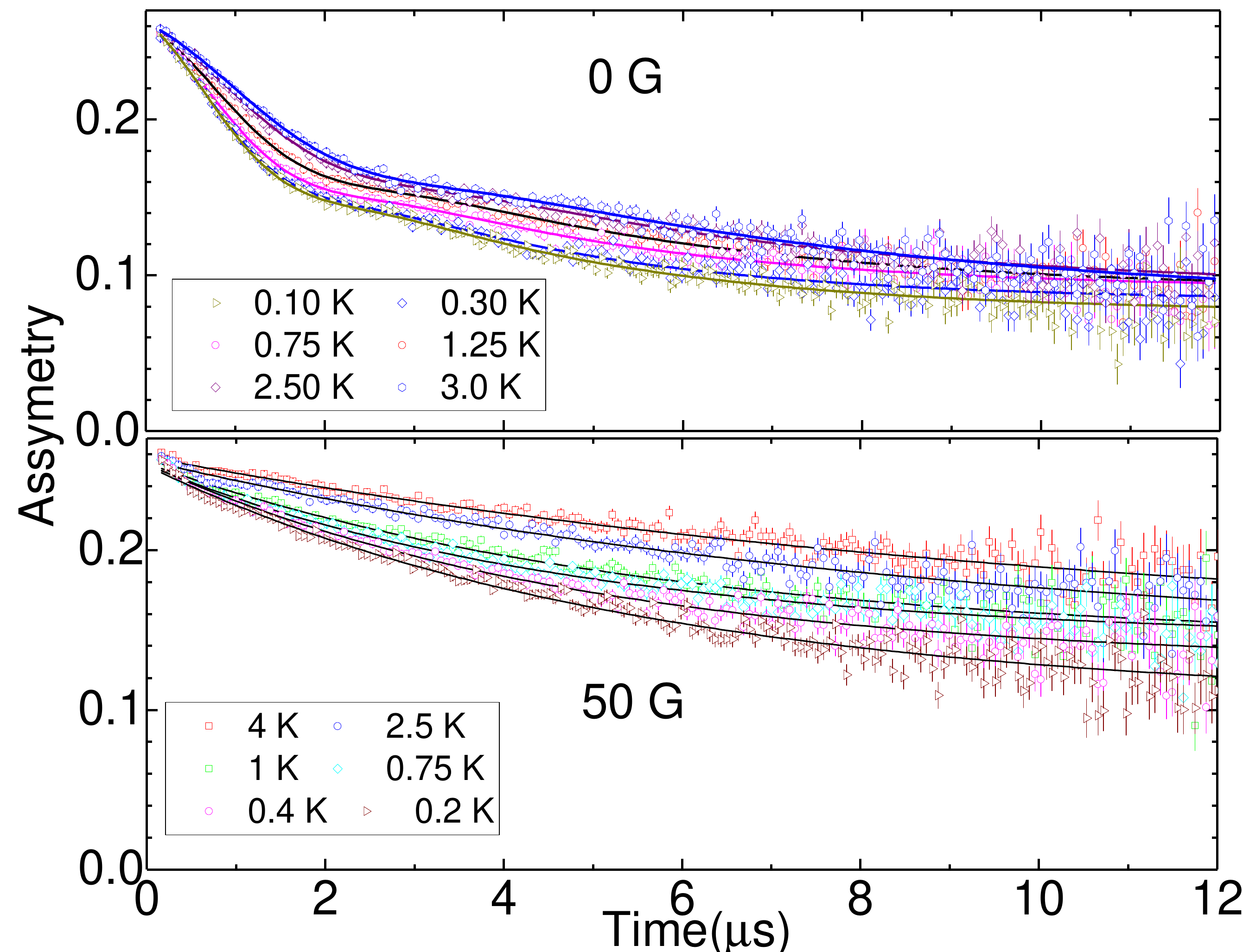}
\caption{\label{fig:ZF-musr-spectra} Top panel: True ZF-\muSR\ time spectra measured at ISIS. Bottom panel: \muSR\ time spectra collected at 50\,G applied longitudinal field.   
Lines indicate the theoretical description as detailed in the text.}
\end{figure}

\textit{Experimental.} Single crystalline and polycrystalline samples of \naybs\ were prepared according to Ref~[\onlinecite{NaYbS2-first-Baenitz}]. \muSR\ experiments were performed at the 
ISIS, UK using the MUSR instruments. For ISIS measurements, 300\,mg of a powder sample, mixed with small amount of GE Varnish to ensure good thermal contact, was dispersed on a silver plate with a radius of 10\,mm. The \muSR\ data were analyzed with the free software packages Mantid.
\textit{\muSR\ results: absence of long range ordering.} Representative ZF and 50\,G (LF)-$\mu$SR asymmetry spectra, measured in a wide temperature ranges, are shown in Fig.~\ref{fig:ZF-musr-spectra}. In general, implanted muons (here positive muons) are highly sensitive to the local magnetic fields with a resolution about ($B = \frac{2\pi}{\gamma_{\mu}} \nu_{\mu}$) $\approx$ 0.1\,mT produced by the adjacent Yb$^{3+}$ spins.~\cite{PhysRevB.95.121111} This makes $\mu$SR as an ideal probe to detect the presence of any tiny static magnetism. It is clear that the present ZF-$\mu$SR spectra do not display any of the characteristic signals originating from static magnetism: 1.) Any spontaneous coherent oscillations in the studied temperature range down to 0.05\,K up to 20\,$\mu$s time range (time spectra are shown upto 12 $\mu$s),  
2.) Strong damping of the muon depolarization 
or 3.) the 1/3 recovery tail of the muon polarization due to random distribution of the static field. On the contrary muon depolarizes faster as cooling. These points demonstrate the absence of a well defined or disordered static magnetic field at the muon stopping site ruling out the possibilities of any long range ordered state of Yb$^{3+}$ moments in \naybs. 
\\
\indent
The ZF-time spectra can be adequately described by the following function in the whole temperature range studied:
\begin{align}
A(t)= A_1G^{\mathrm{KT}}(t, \sigma_{\mathrm{KT}}) + A_2 e^{-\lambda t} + B_{\mathrm{bg}}
 \label{eqn:muonAsymmetry}
\end{align}
where $A_{1}$, $A_{2}$ represents the initial asymmetry, $B_{\mathrm{bg}}$ is the constant background predominantly because of the muons stopped outside the sample. $\sigma_{\mathrm{KT}}$ and $\lambda$ 
are the width of the static field distribution and muon relaxation rate, respectively. To describe the zero-field data adequately two components are needed: One is very small static fraction and the other one is exponential relaxation function, respectively. The former can be easily decoupled by a small amount of longitudinal field, and the later relates to the cooperative spin dynamics of Yb$^{3+}$ spins. Both the contributions, as reflected in $\sigma_{\mathrm{KT}}$ and $\lambda$ increase as lowering the temperature (not shown here). We have also attempted to describe the data using a Dynamic Gaussian Kubo-Toyabe (G$_{\mathrm{Z}}^{\mathrm{DKT}}(t, \sigma_{\mathrm{DKT}}$) and a product function of Gaussian Kubo-Toyabe and exponential decay function (G$_{\mathrm{Z}}^{\mathrm{KT}}(t, \sigma_{\mathrm{KT}})e^{-\lambda t}$). However, neither of these expressions capture the behavior well across the whole temperature range.

\begin{figure}[h]
\includegraphics[width=\columnwidth]{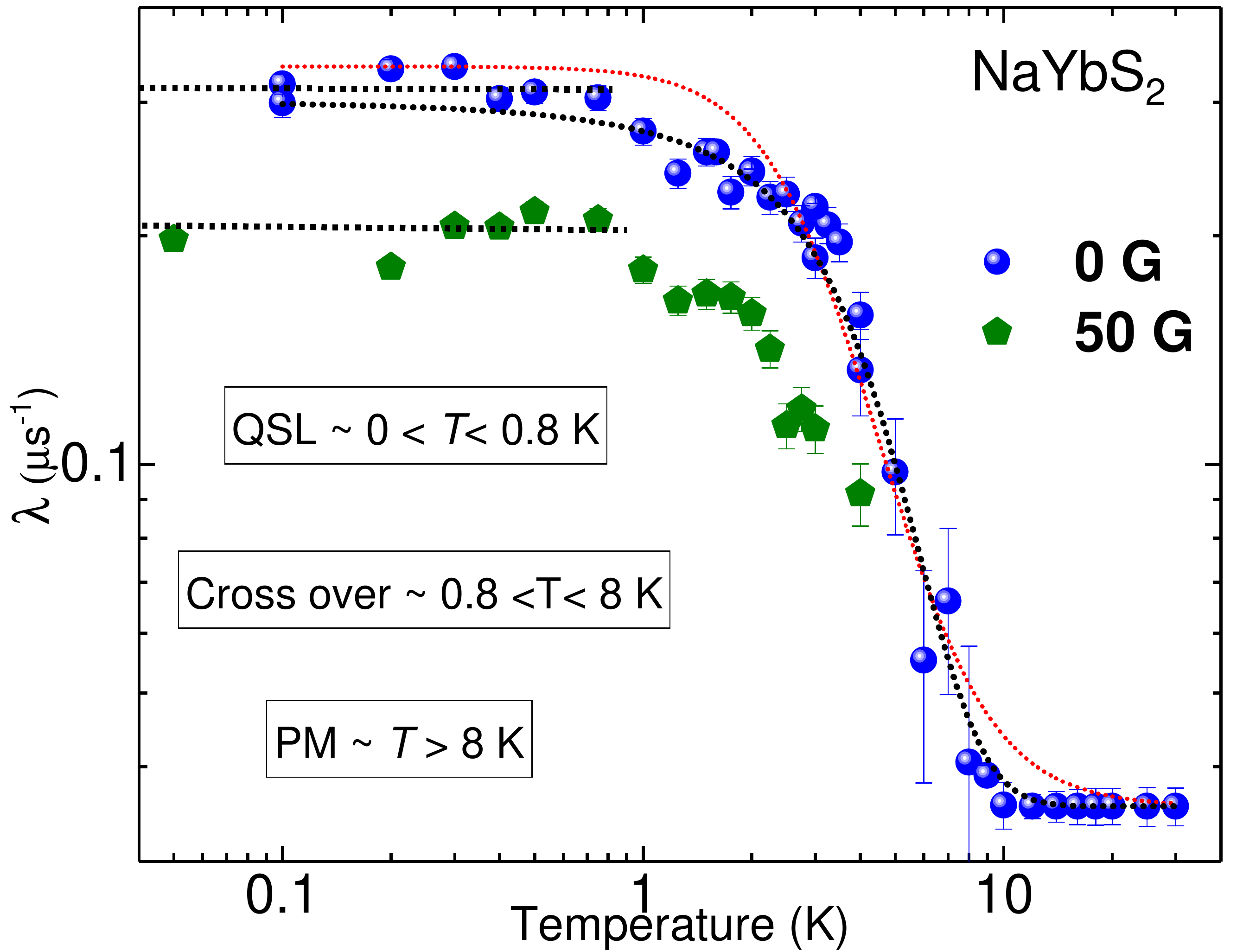}
\caption{\label{fig:relaxation-rate} (Blue spheres and green pentagons) Temperature dependence of the ZF and 50\,G-\muSR\  longitudinal relaxation rates of \naybs. Dotted horizontal lines are guide to the eyes. Lines are theoretical descriptions as detailed in the main text.}
\end{figure}

The bottom panel of Fig.~\ref{fig:ZF-musr-spectra} shows the temperature dependence of $\mu$SR time spectra at 50\,G in LF. It appears that by applying only about 50\,G LF, the static contribution is decoupled, and the $\mu$SR time relaxation spectra follow a single exponential decay: 
A(t)= $A_0 e^{{-\lambda}t}$ + $B_{\mathrm{bg}}$.
 \label{eqn:muonAsymmetry}
The observation of single exponential relaxation depolarization over the whole temperature range investigated evidences that \naybs\ is a dense electronic system reflecting an homogeneous electronic effect. In contrast, the YbMgGaO$_4$ system behaves differently, where $\mu$SR time spectra were adequately described only by using a stretched exponential function, and the stretching parameter varied from 1 to 0.66 while cooling. This further supports the view that \naybs\ is electronically homogeneous.
\\
\indent 
In general, the longitudinal $\mu$ relaxation rate $\lambda$ (=1/$T_1^{\mu}$) can be correlated to spin auto-correlation function by $1/T_1^{\mu} \sim \int^{+\infty}_{0}\left\langle \textbf{S}(t)\textbf{S}(0)\right\rangle \cos(\gamma_{\mu} H_{\mathrm{LF}} t)dt$, where $H_{\mathrm{LF}}=\omega/\gamma_{\mu}$ is the longitudinal applied magnetic field.   
Thus, $\lambda$ depends significantly on the applied external longitudinal magnetic field and on the different correlation functions, $S(t)=\left\langle \textbf{S}(t)\textbf{S}(0)\right\rangle$ for the interacting Yb$^{3+}$ spins. For an exponential correlation function $S(t)\,=\,e^{{-\nu}t}$ leads to the usual Lorentzian spectral density, and this leads $\lambda= 2\Delta^2 \nu/(\nu^2+\gamma_\mu^2H_{LF}^2)$ 
\label{eqn:lambda-field} 
where $\Delta$ is the fluctuating component of the field at the muon site perpendicular to its initial polarization, $\nu$ is the fluctuation frequency, and  $\gamma_\mu$ (=2$\pi\times$ 135.5 MHz/T) is the muon gyromagnetic ratio. The field variation of $\lambda$ may therefore reflect the underlying field distribution rather than field tuned spin dynamics. 

Above $T\sim$10\,K, $\lambda$ (=1/$T_1^{\mu}$) shows an almost temperature independent feature. Considering dimensionality two and a spin coordination number $z$\,=\,6 for Yb$^{3+}$ on triangular lattices, Yb$^{3+}$ spin fluctuation rate in the high $T$ limit can be estimated by using $\nu$\,=\,$\sqrt{z}J_0S/h \sim 1.2\times10^{11}$\,Hz and $\sim 3.5\times10^{11}$\,Hz while $J_{0\parallel}$\,=\,(4.5\,K) and $J_{0\bot}$\,=\,(13.5\,K), respectively.~\cite{{NaYbS2-first-Baenitz},{PhysRevLett.73.3306}} Given that when external field is zero, $\lambda$\,=\,$2\Delta^2/\nu$, gives the internal field distribution 
$\Delta_{\parallel}\,\sim$\,44.8\,$\mu$s$^{-1}$ and $\Delta_{\bot}\sim$77.6\,$\mu$s$^{-1}$ in the high temperature range, which are $<$$<\nu$(1.2$\times10^{11}$\,Hz, 3.5$\times10^{11}$\,Hz). This confirms that the muon spin relaxation is in the fast fluctuation limit.~\cite{{PhysRevLett.93.187201},{PhysRevLett.73.3306},{PhysRevLett.117.097201}}

\textit{\muSR\ rate: collective spin dynamics.} 
Figure~\ref{fig:relaxation-rate} shows the obtained fitting parameter $\lambda$  
as a function of temperature for ZF and for 50\,G. With lowering the temperature $\lambda$ increases constantly, and then saturates to a value of $\lambda_{\mathrm{max}}$ $\approx$ 0.3\,$\mu$s$^{-1}$ below 0.8\,K. 
From the temperature dependence of $\mu$SR investigations, at ZF three exclusive regions can be ascertained. Above 8\,K $\mu$ relaxation rates are temperature independent. This is typical for a paramagnet where spins are short timed correlated. This is in agreement with bulk magnetization data. Below 8\,K, as the temperature goes down (0.8\,K $<T<$ 8\,K) a crossover region is evident, and there $\mu$SR rates enhance dramatically almost 700\,\%, which is nearly one order of magnitude higher than that had been observed in YbMgGaO$_4$. 
With further lowering the temperature below 0.8\,K, $\mu$SR rates saturate to a constant value down to the lowest temperature studied. These sustained spin fluctuations are similar to the other QSL and frustrated magnets~\cite{{PhysRevLett.73.3306},{PhysRevLett.117.097201},{PhysRevLett.92.107204},{PhysRevLett.110.207208}} and, in general, appear to be a signature of QSL ground state. The red dotted line represents the logistic growth function. Clearly this doesn't describe the data adequately. On the other hand the black dotted line is the Boltzman sigmodial/growth function which describe the data much better down to 0.8\,K.  

\begin{figure}[h]
\includegraphics[width=\columnwidth]{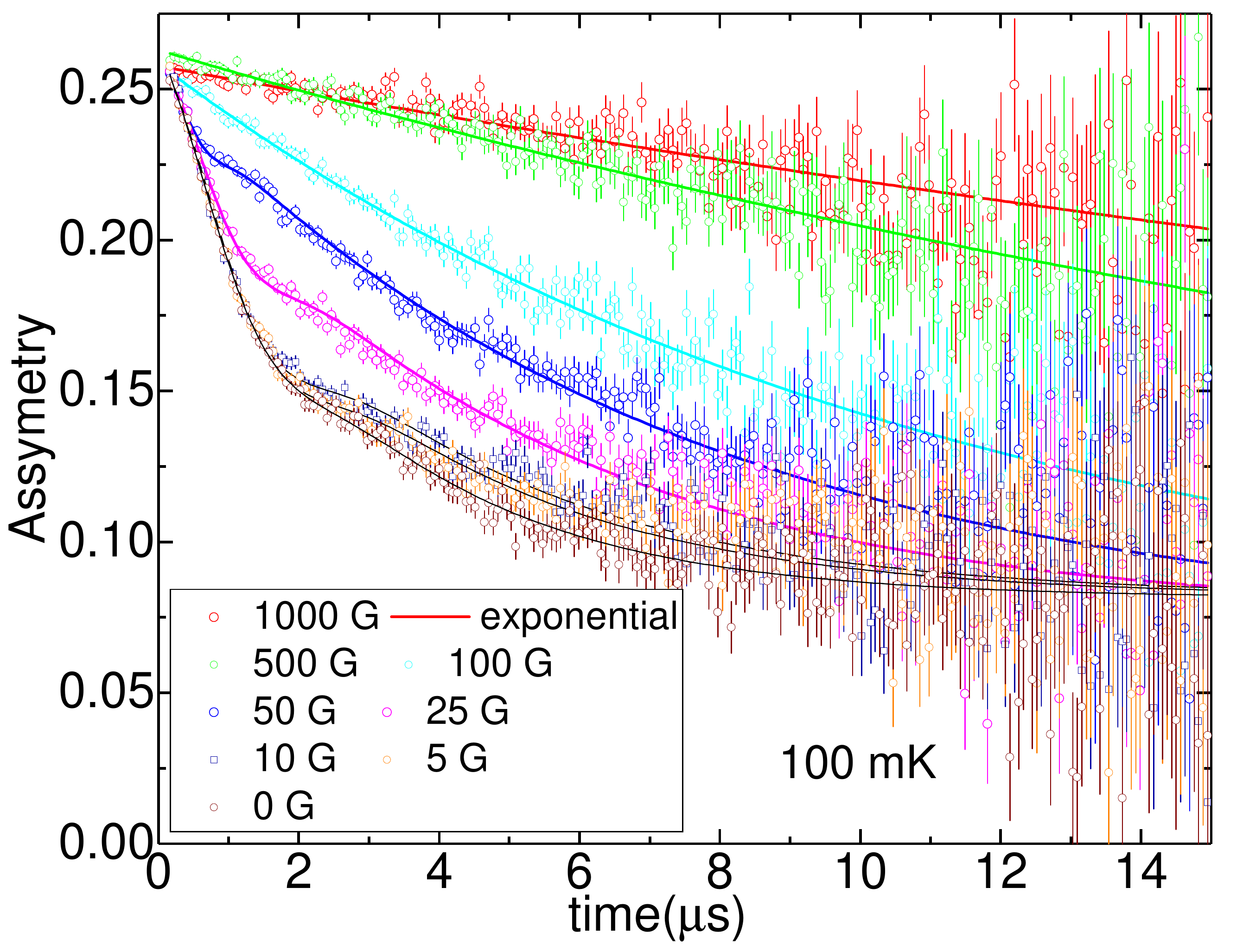}
\caption{\label{fig:LF-spectra} Representative LF-\muSR\  time spectra collected at 100\,mK. Lines are theoretical descriptions as detailed in the main text.}
\end{figure}

Next step is to study the LF effects on $\mu$ polarization to probe the nature of the spin dynamics in \naybs. It is generally accepted that when $\mu$ depolarizes because of the presence of static field distribution, a longitudinal field greater than the static field immediately decouples the $\mu$ depolarization. In contrast, when $\mu$ needs large LF to decouple, then this is most likely the effect of only fluctuating spins. Figure~\ref{fig:LF-spectra} shows the representative LF time spectra at 0.1\,K. It  is seen that even 1000\,G LF is not sufficient to completely decouple the $\mu$ polarization at 0.1\,K ($T\rightarrow$\,0). Similar effects were also found at 4\,K i.e. in the crossover regime. In case of \naybs\ noticeably, however, the $\mu$ signal decouples relatively smaller fields in comparison to YbMgGaO$_4$ (S=1/2) or Ba$_3$NiSb$_2$O$_9$ (S=1), despite that in both the compounds not only the magnetic ions are located on the triangular lattice, but also exhibit a similar $\mu$ relaxation rate ($\lambda \sim$ 0.3\,$\mu$s$^{-1}$) values (plateau) as approaching $T\rightarrow$\,0. On the other hand, in NaYbO$_2$ system the residual relaxation rate is $\lambda$ $\sim$ 1\,$\mu$s$^{-1}$ which is much larger than \naybs\ probably because of the larger CEF's.~\cite{ding-2019}  

\begin{figure}[h]
\includegraphics[width=\columnwidth]{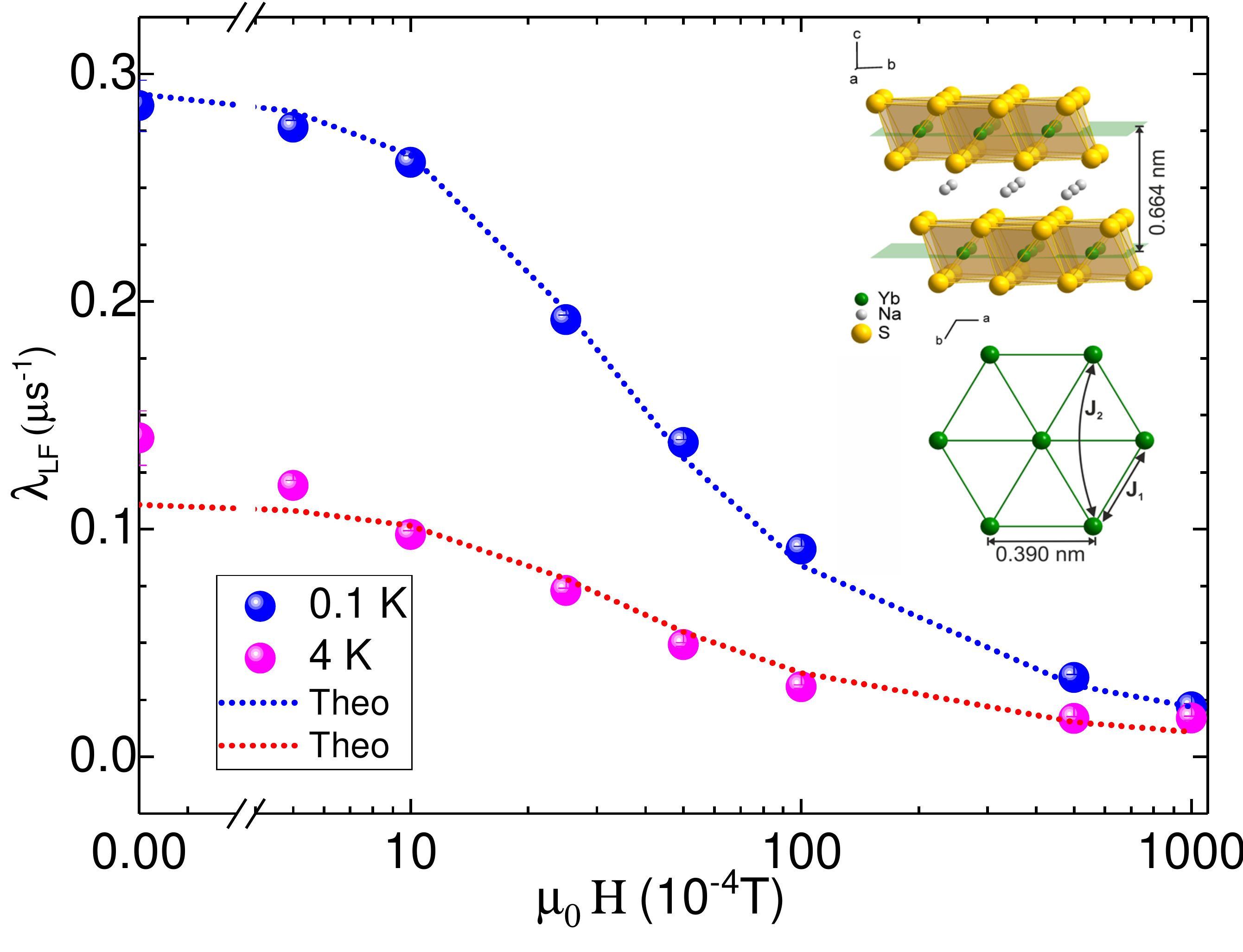}
\caption{\label{fig:LF-rate} (Blue spheres and magenta spheres) Field dependence of the longitudinal relaxation  rate ($\lambda_{LF}$ at 0.1\,K and 4\,K). Inset shows the structure of the \naybs.}
\end{figure}
For YbMgGaO$_4$ (S=1/2), even 1800\,G LF, and for Ba$_3$NiSb$_2$O$_9$ 8000\,G LF, were not sufficient to completely decouple the muon relaxation.~\cite{PhysRevB.93.214432}  
It seems that in \naybs\ the realization of the persistent spin dynamics is rather weak.
Whether or not this is a manifestation of quantum spin liquid ground state or some other new phase is a subject of further investigations. On the other hand it is worthwhile to consider that the existence of very slow fluctuations and/or some slowly freezing of the spins without any transition to a spin glass or a long range ordered value.  
Thus, \naybs\ represents itself as an interesting compound which does not present any sign of ordering. In addition, this gives a demonstrable frustration parameter (4.5/0.05)$>$90. By way of comparison, for NaYbO$_2$ system the frustration parameter value is ($\theta_CW$/0.050 K $>$ 200).~\cite{ding-2019}
\\
The obtained $\lambda_{\mathrm{LF}}$ as a function of field ($\mu_0H_{\mathrm{LF}}$) for two different temperatures are shown in the Fig.~\ref{fig:LF-rate}. In the low temperature limit to describe $\lambda_{\mathrm{LF}}$ the simple exponential correlation function is not enough, the spin dynamic correlation function should take a more general form, $S(t)\sim(\tau/t)^xe^{(-\nu t)}$, where for a  simple exponential case $x$\,=\,0.~\cite{{PhysRevB.64.054403},{PhysRevLett.92.107204},{PhysRevB.84.100401}} The $\mu$ relaxation rate can be represented by the following equation 
$\lambda (H)\,=\, 2\Delta^2\tau^x \int^{\infty}_{0}t^{-x}\exp(-\nu t)\cos(2\pi\mu_0\gamma_\mu Ht)dt$. It is seen that in Fig.~\ref{fig:LF-rate}, $\lambda_{\mathrm{LF}}$ can better be described by the equation with $x\sim0.44$ and $\nu\sim$ 1.7 $\times$10$^6$ Hz. Additionally, at 4\,K, which is a representative temperatures at the cross-over regime, the $\lambda_{\mathrm{LF}}$ appears to be better described $x\sim0.49$ and $\nu\sim$1.7 $\times$10$^6$ Hz. This indicates much slower fluctuations with respect to high temperature paramagnetic state suggesting a long time spin correlations and the  Yb$^{3+}$spins are entangled at low temperatures.
\\
\indent
Contextually NaYbO$_2$ has received significant attention, and  is reported to be another disorder free triangular lattice QSL with a field tunable quantum disorder ground state.~\cite{{Ranjit-2019-PhysRevB.99.180401},{Bordelon-Mitchell-M-2019},{ding-2019}} NaYbO$_2$ system is claimed to be an excellent system for studying quantum disorder ground state in comparison to YbMgGaO$_4$. Very recently we became aware that in external field \naybs\ orders antiferromagnetically starting at 1\,T (Baenitz et al to be published), but in lower fields \naybs\ represents a critical QSL similar to NaYbO$_2$.~\cite{{Ranjit-2019-PhysRevB.99.180401},{Bordelon-Mitchell-M-2019}} 
\\
\indent
The observation of this two component features, in particular the presence of small amount of static component is not new only for \naybs\ system like other delafossites. The recent heat capacity measurements on this compound as reported elsewhere~\cite{NaYbS2-first-Baenitz} showed a maximum at around 0.8\,K. This was conjectured as the emerging spin liquid phase most likely with partially gapped  magnetic excitations. Another likely scenario is that the low temperature spin dynamics are not sufficient to completely subdue magnetic order. The observed peak corresponds to a partial (probably short-range) magnetic order of a minor fraction of spins, whereas the major part remains fluctuating. This scenario can not be completely ruled out considering the two component features of the muon time spectra. In light of this discussion, it is worthwhile to note that the other delafossites NaYbO$_2$ and KCeS$_2$ also show the same two components in their muon spectra~\cite{{ding-2019},{sarkar-2019}}, but they have roughly equal amplitudes. In contrast, the static contribution for NaYbS2 is $<$\,5\,$\%$.
\\
\textit{Conclusions.} 
In conclusion, a detailed \muSR\ study on the \naybs\ system is presented. There is no sign of long range magnetic ordering at least down to 0.05\,K. \muSR\ relaxation rate $\lambda$ values below $\sim$\,0.8\,K are constant suggesting a cooperative quantum disordered ground state in \naybs. Taken all together, that is the low dimensionality, high anisotropy, high frustration index and present \muSR\ studies suggest \naybs\ to be a disorder free triangular lattice which hosts QSL ground state. But what kind of QSL, and what kind of excitations are relevant in \naybs\ demands further theoretical and experimental investigations.    
\\
\textit{Acknowledgments.} This work was financially supported by the Deutsche Forschungsgemeinschaft (DFG) within the SFB 1143 “Correlated Magnetism – From Frustration to Topology”, project-id 247310070 (Projects C02 and B03), GR\,4667/1. We gratefully acknowledge the Science and Technology Facilities Council (STFC) for access to muon beamtime at ISIS. We thank M. Baenitz and D. Inosov for helpful discussions. 

\bibliography{ybnas2}

\end{document}